\documentclass[10pt]{article}

\usepackage[utf8]{inputenc}
\usepackage[english]{babel}
\usepackage[toc,page]{appendix}

\usepackage{amsmath,amsfonts,amssymb,amsthm}
\usepackage{graphicx}
\usepackage{float}
\usepackage{hyperref}
\usepackage{dsfont}
\usepackage{subcaption}
\usepackage[format=plain,font=small,textfont=it]{caption}
\usepackage[dvipsnames]{xcolor}
\usepackage{multirow} 
\usepackage{bookmark}
\usepackage{mathrsfs}

\graphicspath{{img/}}

\usepackage[colorinlistoftodos,prependcaption,textsize=tiny]{todonotes}

\newcommand{\ket}[1]{\left|#1\right\rangle}
\newcommand{\bra}[1]{\left\langle #1\right|}
\newcommand{\ra}{\rangle}

\newcommand{\nn}{\nonumber}

\newcommand{\Wthree}[6]{\left(\begin{array}{ccc} #1 & #2 & #3 \\ #4 & #5 & #6 \end{array}\right)}
\newcommand{\Wfour}[9]{\left(\begin{array}{cccc} #1 & #2 & #3 & #4 \\ #5 & #6 & #7 & #8 \end{array}\right)^{(#9)}}
\newcommand{\Wsix}[6]{\left \{ \begin{array}{ccc} #1 & #2 & #3 \\ #4 & #5 & #6 \end{array}\right \} }

\newcommand{\coolname}{\texttt{sl2cfoam }}

\newcommand{\dtre}{{\Delta_3}}

\begin{document}

\title{Numerical analysis of spin foam dynamics and the flatness problem}

\author{\Large{Pietro Don\`a\footnote{dona@cpt.univ-mrs.fr}, \ } \Large{Francesco Gozzini\footnote{gozzini@cpt.univ-mrs.fr}, \ } \Large{Giorgio Sarno\footnote{sarno@cpt.univ-mrs.fr} \ }
\smallskip \\ 
\small{CPT, Aix-Marseille\,Universit\'e, Universit\'e\,de\,Toulon, CNRS, 13288 Marseille, France}
}

\date{\today}

\maketitle

\begin{abstract}
In this paper we apply a recently proposed numerical algorithm for finding stationary phase points in spin foam amplitudes. We study a spin foam amplitude with three vertices and a bulk face in 4D BF theory. We fix the boundary coherent states to three possible triangulations, one with zero deficit angle on the bulk face and two with non-zero deficit angle. We compute the amplitude numerically and we find a stationary phase point already at low spins in all the three cases. 
We comment on how this result contrasts with the claims of flatness problem in spin foam theories.
We point out where these arguments may be misleading and we propose further computations to definitively answer the question.
\end{abstract}

\section{Introduction and Motivations}
\label{sec:intro}

The EPRL-FK spin foam theory \cite{Engle:2007wy,Freidel:2007py} is an attempt to define the dynamics of loop quantum gravity. It provides a regularized, background-independent and Lorentz covariant quantum gravity path integral with the definition of a partition function on a triangulation. The theory assigns transition amplitudes to spin network states on the boundary of a triangulation.

Recovering General Relativity (or at least a discrete version of it) in the semiclassical limit is a fundamental test for spin foam theories. 
The large spin limit of the EPRL vertex amplitude has been largely explored in both euclidean \cite{Barrett:2009gg} and lorentzian versions \cite{Barrett:2010}. Remarkably, under a homogeneous rescaling of all the spins, the single vertex amplitude with coherent boundary data contains the Regge action, a discrete formulation of General Relativity.

However, in the case of extended triangulations, the semiclassical limit of the theory is not completely understood. In particular, the question of what semiclassical geometries dominate the summations over bulk degrees of freedom is still open.

Different calculations suggest that the summation over bulk degrees of freedom is dominated by flat geometries\footnote{Precisely, only geometries with a deficit angles around bulk faces multiple of $4\pi/\gamma$.}. This observation, often called \emph{flatness problem}, has been mentioned by Freidel and Conrady \cite{Conrady:2008mk}, Bonzom \cite{Bonzom:2009hw}, and Han \cite{Han:2013ina}. Hellmann and Kami\'nski \cite{Hellmann:2013gva} proposed a different analysis, based on the calculation of the wavefront set of the spin foam partition function and a geometrical interpretation of its variables, and they found a similar problem.  

A more refined stationary phase analysis of the euclidean EPRL model has been carried out by Oliveira in \cite{Oliveira:2017osu}. They include all the constraints on the boundary data necessary to obtain a non-exponentially suppressed vertex amplitude. The initial claim that the flatness problem is not present in that setting has been recently questioned by Kaminski and Engle \cite{EnglePrivateComm}. It has been suggested that taking only large spins is not the right way to access the semiclassical regime of the theory \cite{Magliaro:2011zz,Han:2013hna, Asante:2020qpa}. 

The importance of numerical methods for the investigation of spin foam theories grew considerably in recent years. The software library \coolname is a C based high-performance library and is the foundation for the numerical computation of general spinfoam amplitudes, both in 3 and 4 dimensions, with topological BF or lorentzian EPRL models \cite{Dona:2018nev}. The library is based on the factorization \cite{Speziale:2016axj} and was used to explore the large spin limit of both models \cite{Dona:2017dvf,Dona:2019dkf}.

While still at the early stages, the evaluation of many-vertices spinfoam amplitudes is possible. In a recent work \cite{Gozzini:2019kui}, we have developed an algorithm to determine the existence and estimate the position of stationary phase points in the summations over the spins of internal faces. We tested its effectiveness applying it to BF spin foam theory in 3D (the Ponzano-Regge model), where the stationary phase points are directly connected to the solution of the equations of motion of euclidean Regge calculus.  We adapt the algorithm to the BF spin foam theory in 4D. This topological model is the starting point for the construction of physical spin foam models like the EPRL model. For this reason, there are many similarities between the two. For example, the boundary states of the transition amplitudes are the same, as well as the semiclassical geometries emerging from the single vertex asymptotics. What geometries dominate the BF transition amplitudes in the large spin limit is an interesting question on its own. 

In this work, we focus on the $\dtre$ triangulation, formed by three 4-simplices sharing a common triangle. It is the simplest triangulation with many vertices and a bulk face, and it is the standard example used in the flatness problem literature \cite{Collet:2016doe,Oliveira:2017osu,Hellmann:2013gva}. 

Solutions of the classical equation of motion dominate the summation over the bulk degrees of freedom in the large spin limit. We expect them to manifest through the presence of stationary phase points. Note that in classical Regge calculus, the $\dtre$ triangulation is too simple to give non-trivial dynamics since boundary conditions fix all the lengths. Hence, this triangulation is not a good example to probe the large spin limit of the EPRL model if we suppose the classical underlying theory to be (area-angle) Regge Calculus. Nevertheless, the study of this amplitude can teach us some valuable lessons.

We find an unexpected result. We construct three examples of the 4D euclidean Regge $\dtre$ triangulation. One is the flat triangulation characterized by a zero deficit angle around the shared triangle. The other two are curved triangulations. We compute the coherent amplitude associated with each triangulation, and we look for stationary phase points in the internal spin, dual to the shared triangle. We find the presence of a stationary phase point in all the cases. We estimate its value, and we find it compatible with the area of the dual triangle of the prescribed triangulation.

If we apply arguments similar to the ones claiming flatness of the EPRL model to BF theory, we would infer that curved triangulations should be suppressed in the sum. Therefore we would not expect to find any stationary phase point in this case. This shows tension between our results and the arguments declaring that the EPRL model is flat.

The  paper  is  organized  as  follows. In Section \ref{sec:theory} we review the spin foam formulation of BF theory and the $\dtre$ transition amplitude. Section \ref{sec:d3geometry} describes the Regge triangulations we use to construct the boundary data of the amplitude. In Section \ref{sec:numerical} we illustrate the algorithm we use to search for stationary phase points and its application. We review the main arguments of the flatness problem of the EPRL model applied to the topological BF model in Section \ref{sec:flatness}. We conclude with a summary of our numerical results and their implications.

All the computations are performed on the CPT servers in Marseille. Each machine is equipped with a 32 cores CPU Intel(R) Xeon(R) Gold 6130 with a base frequency of 2.10 GHz and 196 GB of RAM. In the public repository \cite{repo} we publish the C code used to calculate the transition amplitudes along with the Wolfram Mathematica notebooks used to prepare the boundary data and to perform the numerical stationary phase point analysis.


\section{BF theory and the $\dtre$ transition amplitude}
\label{sec:theory}
General Relativity in four dimensions can be written as a BF theory with constraints. This is the starting point in the construction of many spin foam models, as the lorentzian EPRL model. In this paper, we study the spin foam formulation of the $SU(2)$ BF theory on a four-dimensional manifold. For a detailed overview of BF theory and its relations with spin foam models, see \cite{Baez:1999sr} or \cite{Perez:2012wv}. 
A triangulation $\Delta$ of the manifold is dual to a two-complex $\Delta^\star$ that consists of a set of vertices (dual to the 4-simplices), edges (dual to the tetrahedra) and faces (dual to the triangles). We assign an SU(2) holonomy $g_e$ to each half-edge of the triangulation. The partition function of the theory discretized on $\Delta$ is given by 
\begin{equation}
  \mathcal{Z}(\Delta) = \int \prod_e dg_e \prod_f \delta(g_{e_1} \dots g_{e_n}) \ , \qquad \text{ with } e_i \subset f \ ,
\end{equation}
the product $g_{e_1} \dots g_{e_n}$ is the holonomy around the face $f$, and $dg_e$ is the Haar measure on $SU(2)$. We expand the delta function using the Peter-Weyl theorem as $\delta(g) = \sum_j (2j+1) Tr(D^j(g))$, where $D^j(g)$ are the Wigner matrices of the SU(2) representation of spin $j$. We get
\begin{align}
  \label{eq:graphical}
      \mathcal{Z}(\Delta) &= \sum_{j_f} \int \prod_e dg_e \prod_f (2 j_f+1) Tr (D^{j_f}(g_{e_1} \dots g_{e_n})) \\
      & = \sum_{j_f}  \prod_f (2 j_f +1)  \prod_v \raisebox{-10mm}{ \includegraphics[width=4.5cm]{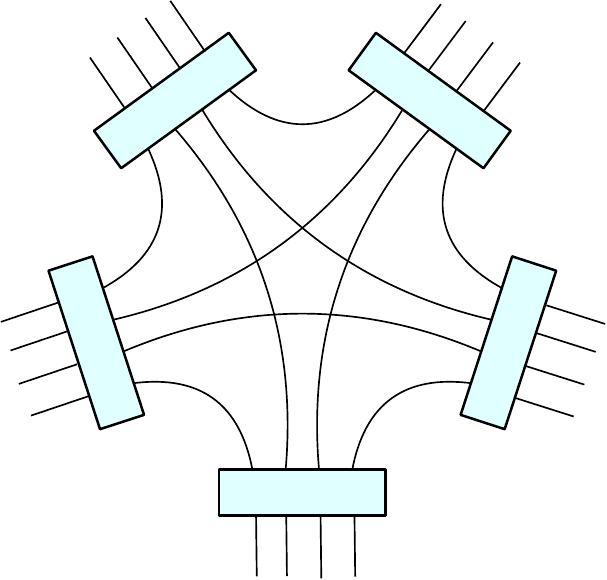}} \ .
  \end{align}
It is useful to introduce a graphical notation for spin foam vertices. Every box represents one integral over SU(2), each line is an irreducible representation of SU(2), and each edge contains four lines. For more details on graphical notations see \cite{Varshalovich1988,Dona:2017dvf}. We reorganized the partition function as a product over the vertices of $\Delta^\star$ (4-simplices of $\Delta$) and a sum over all the spins. 
Furthermore, the integrations over SU(2) can be performed explicitly, introducing a sum per edge over a 4-valent intertwiner space $i_e$. We obtain the familiar form for the partition function in terms of amplitudes
\begin{equation}
  \mathcal{Z}(\Delta)  = \sum_{j_f, i_e}  \prod_f A_f \prod_e A_e \prod_v A_v\ .
\end{equation}
The face amplitude and the edge amplitude are given respectively by the dimensional factors $A_f = 2j_f +1$ and $A_e=2 i_e +1$. The vertex amplitude $A_v$ is given by a $\{15j\}$ symbol. In this work we use the $\{15j\}$ symbols of the first type (see Appendix \ref{app:conventions} for our conventions). 

Curiously, the vertex amplitude of this theory with coherent boundary data contains the Regge action of a 4-simplex in the large spin limit. This connection with discrete gravity has been extensively studied both analytically \cite{Barrett:2009as} and numerically \cite{Dona:2017dvf}, and is  limited to a single vertex. The topological nature of the model is encoded in the delta functions that appear in triangulations with internal faces.  

\begin{figure}[H]
  \centering
  \begin{subfigure}[c]{0.30\textwidth}
  \includegraphics[width=\textwidth]{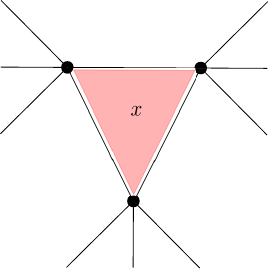}
  \end{subfigure}
  \hspace{0.09\textwidth}
  \begin{subfigure}[c]{0.59\textwidth}
      \includegraphics[width=8cm]{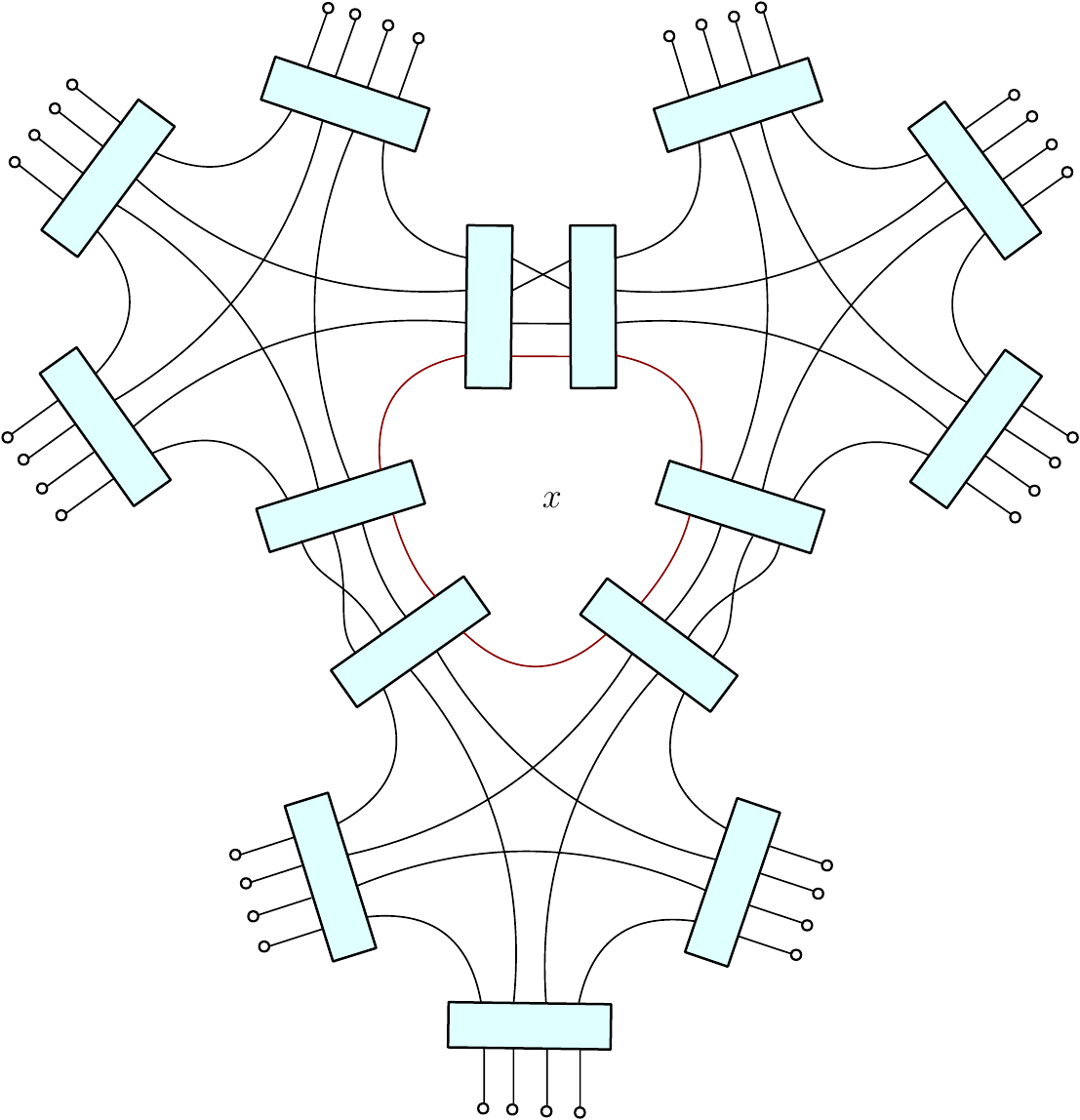}
  \end{subfigure}
  \caption{Left: The 2-complex of the $\dtre$ triangulation. We highlight in red the bulk face.  Right: The $\dtre$ spin foam amplitude in graphical notation. We are omitting a summation over the internal spin $x$ and a dimensional factor $2x+1$. The boxes represent integrals over the $SU(2)$ group, and the strands are irreducible representations labeled by a spin $j_{abc}$. The empty dots on the boundary represents the SU(2) coherent states.}
  \label{fig:spinfoam}
\end{figure}

We study the spin foam transition amplitude of the $\dtre$ triangulation. The 2-complex dual to the $\dtre$ triangulation is represented in Figure \ref{fig:spinfoam}. It is the simplest amplitude with 3 vertices, 18 boundary faces and one internal face. Each vertex shares an edge with the other two and has three boundary edges. 

The boundary data consist of coherent states parametrized by 18 spins and 36 unit normals, one spin and two normals per each strand of the spin foam diagram. We denote with $j_{abc}$ each boundary spin and with $x$ the internal spin. In the next section, we study the four-dimensional geometry we use to specify the boundary data; the notation will be clearer then. The spin foam transition amplitude, omitting the summation over $x$ and the associated face amplitude, is represented using the graphical notation in Figure \ref{fig:spinfoam}. We refer to \eqref{eq:d3ampfull} in appendix \ref{app:derivation} for the complete picture with all the spin labels and orientations of the faces. 
We perform the SU(2) integrals and obtain the $\dtre$ spin foam amplitude in terms of $\{ 15j\}$ symbols (see appendix \ref{app:derivation} for more details)
\begin{align}
  \label{eq:d3ampl}
  W_{\dtre} (j_f, \vec{n}_{fa}) & =(-1)^{\chi} \sum_x w_\dtre(x,j_f,\vec{n}_{fa}) \\
  &= (-1)^{\chi} \sum_x (-1)^x(2 x +1) \sum_{k_b} \left( \prod_b (-1)^{k_b}(2 k_b + 1)\right) \sum_{i_a} \left( \prod_a (2 i_a + 1) c_{i_a}(\vec{n}_{fa}) \right) \nn \\
    & \times \left \{
\begin{array}{ccccc} i_1 & j_{345} & k_1 & j_{145} & i_2 \\  j_{235} & x & j_{134} & j_{124} &
j_{245} \\ j_{125} &
k_2 & j_{123} & i_3 & j_{234}  \end{array}\right \} \left \{ \begin{array}{ccccc} i_4 &
j_{123} & k_2 & j_{235} & i_5 \\  j_{136} & x & j_{125} & j_{256} & j_{236} \\ j_{356} & k_3 & j_{156} & i_6 &
j_{126}  \end{array}\right
\} \nn \\ & 
\times \left \{ \begin{array}{ccccc} i_7 & j_{156} & k_3 &
j_{136} & i_8 \\  j_{145} & x & j_{356} & j_{346} & j_{146} \\ j_{134} & k_1 & j_{345} & i_9 & j_{456}  \end{array}\right \} \nn 
\end{align}
where $\chi = 2 (j_{123} + j_{234}+ j_{124}+j_{134}+j_{456}+j_{156}+j_{346}+j_{356}+j_{235}) + j_{123} + j_{345} +j_{156}$ is a phase function only of the boundary spins, $i_a$ are the boundary intertwiners, $k_b$ are the internal intertwiners and the sum over the spin $x$ is bounded by triangular inequalities. The complex coefficients $c_{i_a} (\vec{n}_{fa})$ are the Livine-Speziale coherent intertwiners in the recoupling basis, they depend on the normals $\vec{n}_{fa}$ and areas of the tetrahedron $a$ (see appendix \ref{app:coherentstates} for their definition).


\section{The $\dtre$ geometry}
\label{sec:d3geometry}

The $\dtre$ triangulation is formed by three 4-simplices, sharing a common triangle. The boundary of the triangulation consists of nine tetrahedra and the bulk consists of three tetrahedra all sharing the common triangle $x$.
\begin{figure}[H]
  \centering
  \includegraphics[width=0.9\textwidth]{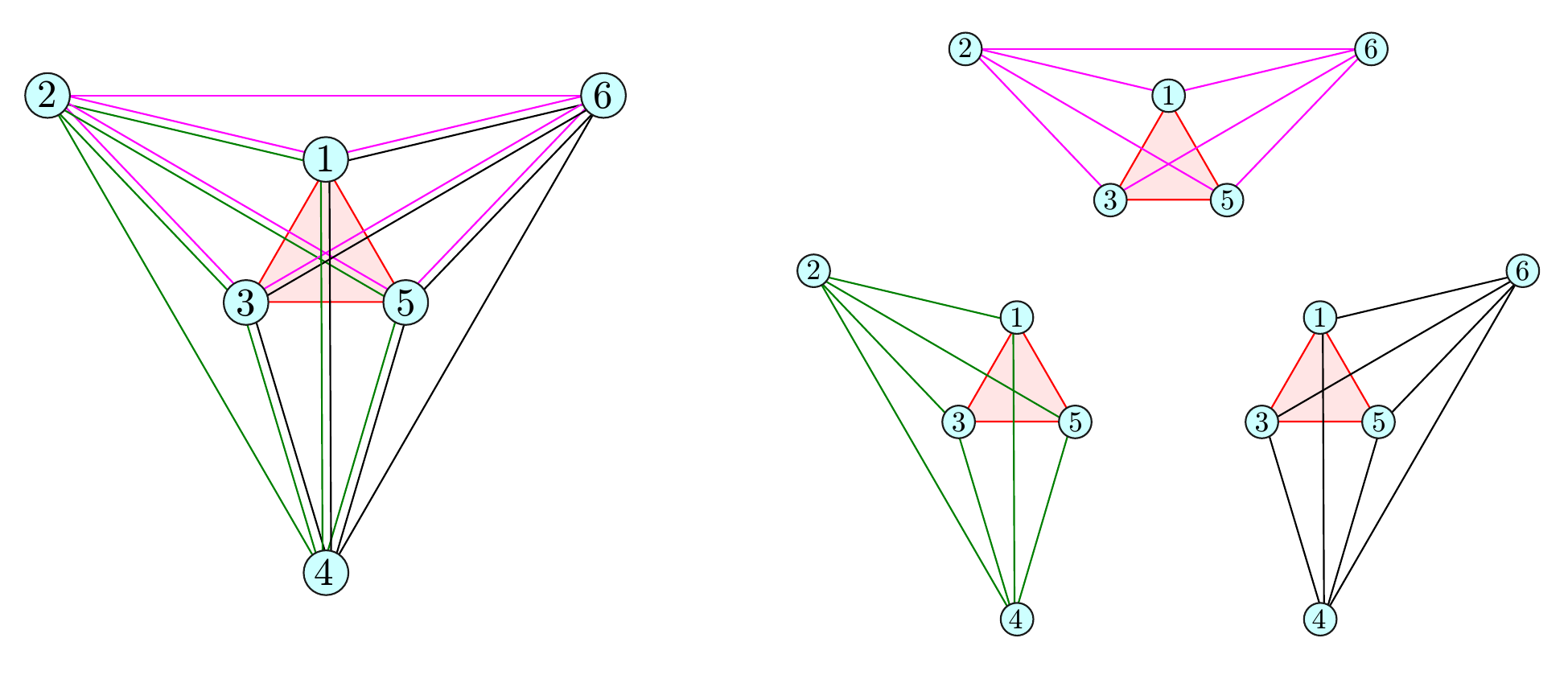}
  \caption{Left: The geometry of the $\dtre$ triangulation. The numbered circles correspond to points. Lines correspond to segments. Each color corresponds to a different 4-simplex. The bulk triangle is highlighted in red. Right: The three 4-simplices are shown separately.}
  \label{fig:geometry}
\end{figure}
We associate to each boundary tetrahedron a coherent state characterized by 4 spins, the areas of its faces, and 4 unitary vectors, the outward normals to the faces of the tetrahedron. 
The triangulation is made of points, segments, faces and tetrahedra. We use the triangulation representation in Figure \ref{fig:geometry} as reference. The boundary spins $j_{abc}$ are labeled by a triple of points $a,b,c$ since they correspond to triangles. We decide to focus on Regge geometries, fully characterized by their 15 lengths $\ell_{ab}$ of the segment joining points $a,b$. Notice that all the lengths belong to at least one boundary tetrahedron, therefore fixing the boundary determines the triangulation completely. 

For simplicity and easier computability, we restrict our analysis to geometries with a high degree of symmetry. We require ``cylindrical'' symmetry forcing the three 4-simplices to be identical. This symmetry requires the bulk triangle to be equilateral ($\ell_{13}=\ell_{35}=\ell_{15}$) and the bulk tetrahedra to be isosceles, having an equilateral base and three identical isosceles faces ($\ell_{12}=\ell_{25}=\ell_{23}=\ell_{16}=\ell_{36}=\ell_{56}=\ell_{34}=\ell_{14}=\ell_{45}$). The remaining three lengths also have to be equal ($\ell_{26}=\ell_{24}=\ell_{46}$). 

Furthermore, we require all the areas of the boundary triangles to be the same, and this fixes $\ell_{13}=\ell_{26}$. The symmetry reduces the degrees of freedom of the Regge triangulation from 15 to 2, for example, $\ell_{12}$ and $\ell_{13}$. In terms of the lengths, we can compute all the geometric quantities of the triangulations: areas, three and four-dimensional volumes and two, three and four-dimensional dihedral angles. It is interesting to compute the area of the boundary triangles $\lambda$ and $\alpha$ the 4D dihedral angle around the bulk triangle of one 4-simplex:
\begin{equation}
\lambda = \frac{1}{4} \ell_{13} \sqrt{4 \ell_{12}^2 -\ell_{13}^2} \ , \qquad \sin^2 \alpha = \frac{3}{4} \ell_{13}^2 \frac{ 12 \ell_{12}^2-7 \ell_{13}^2}{ \left(\ell_{13}^2-3 \ell_{12}^2\right)^2} \ .
\end{equation} 
They are independent variables and under a linear rescaling of the lengths $\lambda$ scales quadratically while $\alpha$ does not scale. In the following, we identify with $\lambda$ the scale of the triangulation. 

Another interesting geometrical quantity for our analysis is the value of the area of the shared triangle
\begin{equation}
\label{eq:geoarea}
x_g = \frac{\sqrt{3}}{4} \ell_{13}^2 = 3 \left(\frac{6+\sin ^2 \alpha +6 \sqrt{1-\sin ^2 \alpha}}{48+\sin ^2 \alpha} \right)^{\frac{1}{2}}\ \lambda \ .
\end{equation}
For Regge geometries curvature can be expressed in terms of deficit angles associated to triangles shared between 4-simplices. In the particular class of symmetric triangulations we are considering the deficit angle associated with the shared triangle is 
\begin{equation}
\delta = 2\pi - 3\alpha \ . 
\end{equation}
We parametrize the triangulations in this paper with $\lambda$ and $\delta$ instead of the two lengths. Once the triangulation is specified, we compute all the normals of the boundary tetrahedra to use them in defining the boundary coherent states. 

In this work we consider the flat triangulation characterized by $\delta=0$ at various scales $\lambda$, a first curved triangulation characterized by $\delta\approx 3.60$ at scale $\lambda=20$, and a second curved triangulation characterized by $\delta\approx 2.47$ at scale $\lambda=20$. The chosen parameters of the curved triangulations are particularly convenient for this numerical analysis. We motivate our choice of focusing on a single scale for the curved configurations in the next section. All the notebooks we used to compute all the geometric data and the boundary normals for the three triangulations are available in \cite{repo}.


\section{Numerical analysis}
\label{sec:numerical}

The algorithm in \cite{Gozzini:2019kui} is designed to determine the presence and estimate the value of stationary phase points in summations over spin foam bulk spins. We adapt it to the amplitude in examination that we study for different choices of boundary data. In this section, we briefly recall features of the algorithm, and we highlight the differences with the original version. We refer to the original paper for a detailed explanation.

The amplitude \eqref{eq:d3ampl} involves a sum over the spin $x$ of the internal face and the intertwiners $k_b$ of the three internal edges. For brevity let us omit the dependence on the boundary data $j_f$, $\vec{n}_{fa}$ on the summand $w_\dtre(x) \equiv w_\dtre(x,j_f,\vec{n}_{fa})$. 
First, we compute $w_\dtre(x)$ for each value of the internal spin $x$ using the library \coolname \cite{Dona:2018nev}. We focus on two quantities 
\begin{equation}
P_w(x) = \sum_{x'\,=\, x_\text{min}}^x w_\dtre(x^\prime) \ , \qquad \text{and} \qquad R_w(x) = \sum_{x' \in I_x^c} w_\dtre(x') \ ,
\end{equation}
where $I_x^{c}$ is an interval centered in $x$ with width $2c$. We call them respectively the \emph{partial sum} and the \emph{running sum}. The algorithm is based on the behavior of these two quantities in the presence of stationary phase points for $W_\dtre$. 

Suppose that $x_0$ is a stationary phase point for $W_\dtre$. The sum over the internal spin interferes destructively away from $x_0$ and constructively near $x_0$. The partial sum $P_w(x)$ stays roughly constant away from $x_0$ and has a sudden ``jump'' near $x_0$. Similarly, the running sum $R_w(x)$ is close to zero away from $x_0$ and peaks at $x_0$. The width of the peak depends on the parameter $c$, the size of the interval that characterizes the running sum.

Second, we consider $\overline{R_w(x)} = |R_w(x)_{c_1}R_w(x)_{c_2}R_w(x)_{c_3}|$ correlating three running sums with different interval sizes $c_i$ and we use a Mathematica's built-in function to locate the peaks of $\overline{R_w(x)}$. We repeat this step for all the possible triples of $c_i$ in a reference set. Since we are working at smaller spins then \cite{Gozzini:2019kui} in this work we use as reference set $c=\sqrt{x_{max}-x_{min}}/2$ plus or minus 50\%. The numerical estimate of the stationary phase point is given by the average of the outcomes of this procedure with an error given by the standard deviation (both values are rounded to the nearest integer). This aims to eliminate the dependence of our analysis from the choice of a particular interval.

\begin{figure}[H]
  \centering
  \begin{subfigure}[c]{0.48\textwidth}
  \includegraphics[width=\textwidth]{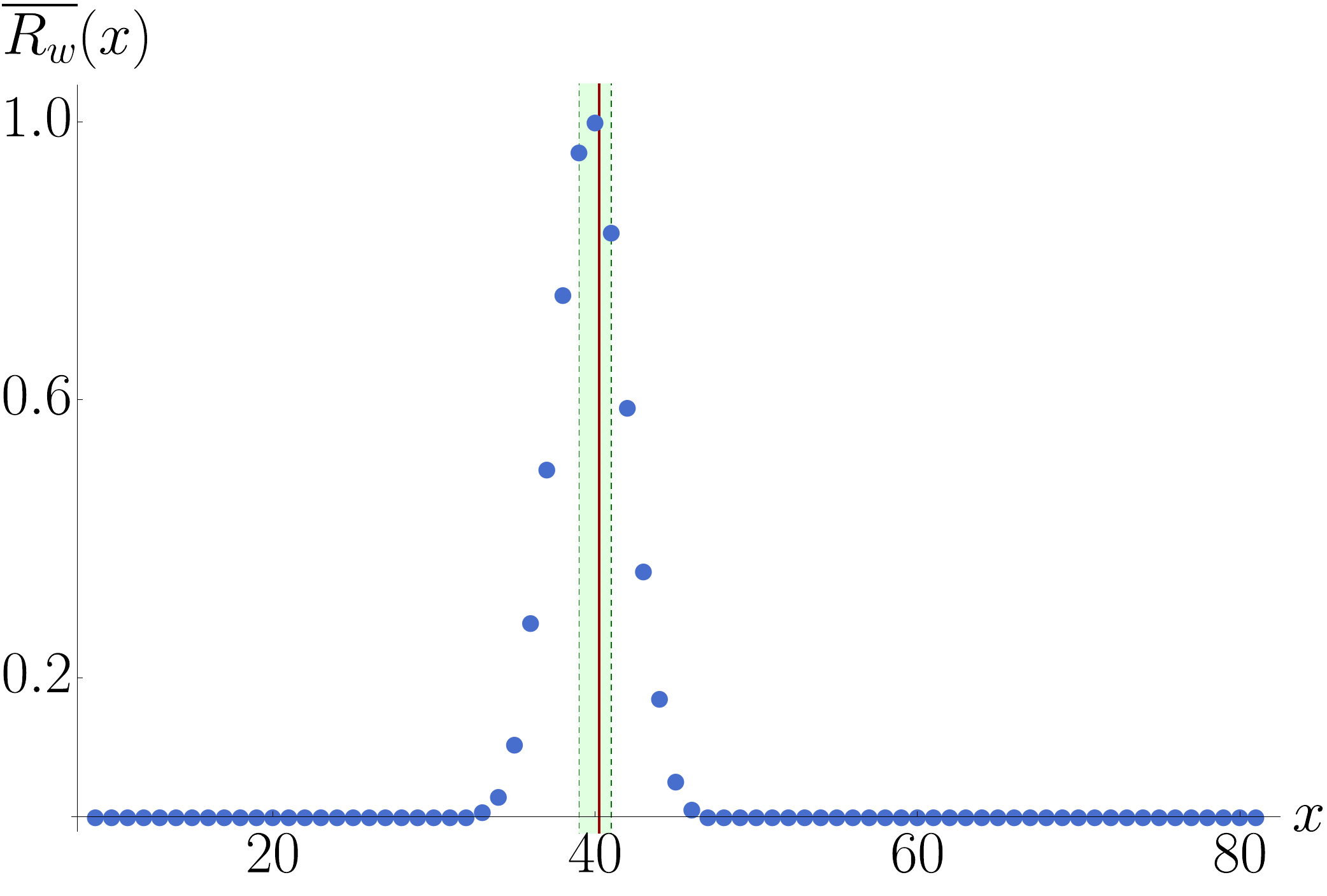}
  \end{subfigure}
  \hspace{0.01\textwidth}
  \begin{subfigure}[c]{0.48\textwidth}
      \includegraphics[width=\textwidth]{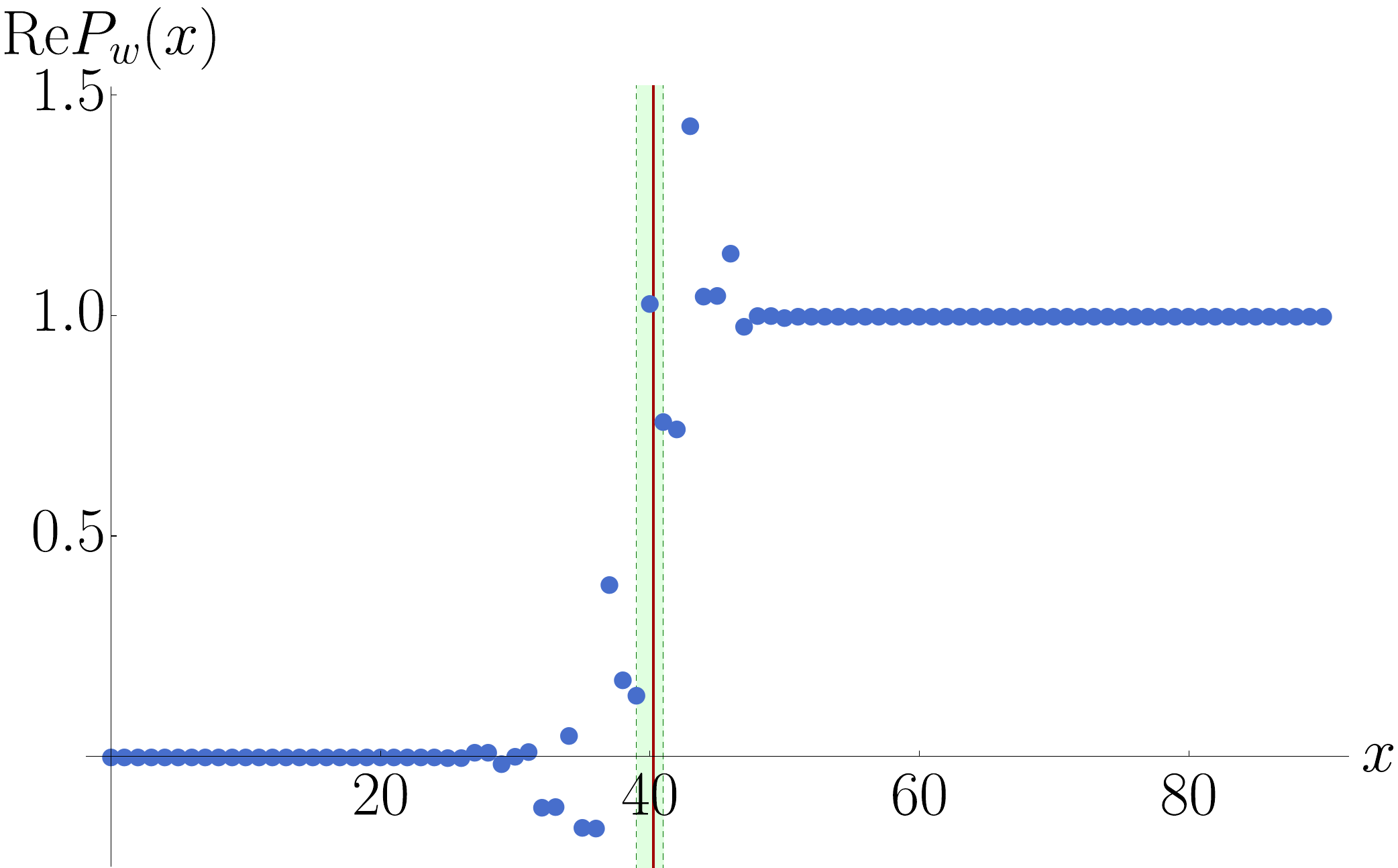}
  \end{subfigure}
  \caption{Flat configuration. Left: Averaged running sum $\overline{R_w(x)}$ with interval sizes $2$,$6$, and $10$. We normalize respect to the value of its maximum. Right: Real part of the partial sum $P_w(x)$ normalized respect the absolute value of the amplitude. In both plots, we mark with a red line the value of the geometrical area of the bulk triangle, and with a green band the values of the numerical estimate of the stationary phase point within one standard deviation. }
  \label{fig:flat}
\end{figure}

The first triangulation we study is the flat one. We report in Figure \ref{fig:flat} the running sum and the partial sum for this configuration at $\lambda = 30$.
Our algorithm is implemented in Mathematica, the notebooks are available in \cite{repo}. We estimate the presence of a stationary phase point for spin $x_{n_0}= 40 \pm 1$. The value of the geometric area shared by the three 4-simplices \eqref{eq:geoarea} for the flat configuration is 
$x_{g_0} = 1.34\ \lambda  \approx 40.2$. It is in perfect agreement with our numerical analysis. 

To corroborate the result we repeat the analysis at other scales $\lambda =10, 12,14,16,18,20$. We summarize in Table \ref{table:flat-lambda} the numerical estimate of the stationary phase point and the geometric area of the shared triangle. Signatures of the stationary phase point are present at all scales, and its position is always compatible with its geometric counterpart. We also note the resources necessary to perform the computations on our server. 

\begin{center}
  \begin{tabular}{c|c|c|c|c}
    & numerical & analytical & memory (GB) & time (h)\tabularnewline
  \hline 
  $\lambda=10:$ &  $13\pm1$ & $13.4$ & $2$ & $0.2$  \tabularnewline
  \hline 
  $\lambda=12:$ &  $16\pm1$ & $16.1$ & $6$ & $0.4$  \tabularnewline
  \hline 
  $\lambda=14:$ &  $18\pm1$ & $18.8$ & $16$ & $0.9$   \tabularnewline
  \hline 
  $\lambda=16:$ &  $21\pm1$ & $21.5$ & $33$ & $3$   \tabularnewline
  \hline 
  $\lambda=18:$ &  $23\pm1$ & $24.1$ & $65$ & $7.8$   \tabularnewline
  \hline 
  $\lambda=20:$ &  $26\pm1$ & $26.8$ & $121$ & $21$  \tabularnewline
  \hline 
  $\lambda=30:$ &  $40\pm1$ &  $40.2$ & $\sim 2500$ & $\sim 2700$  \tabularnewline
  \end{tabular}
  \captionof{table}{    \label{table:flat-lambda} We compare the numerical estimate of the stationary phase point with the geometrical area at different scales. For each calculation, we report the time (in hours) and the memory (in GB) required for these computations on our server. Time and memory for $\lambda = 30$ are an estimation based on the previous points. The computation, in this case, was performed on multiple machines, a piece at a time and took more than two months to complete. }
\end{center}

Notice that the agreement is excellent even at small scales. This confirms the result of \cite{Dona:2017dvf,Dona:2019dkf}, the asymptotic regime is reached already at relatively low spins. Moreover, at larger scales $\lambda$, the computation becomes computationally demanding both on memory and time. This motivated us to focus only on $\lambda=20$ for other boundary data, the best compromise in terms of results and our computational resources.  

\begin{figure}[H]
  \centering
  \begin{subfigure}[c]{4cm}
    \includegraphics[width=7cm]{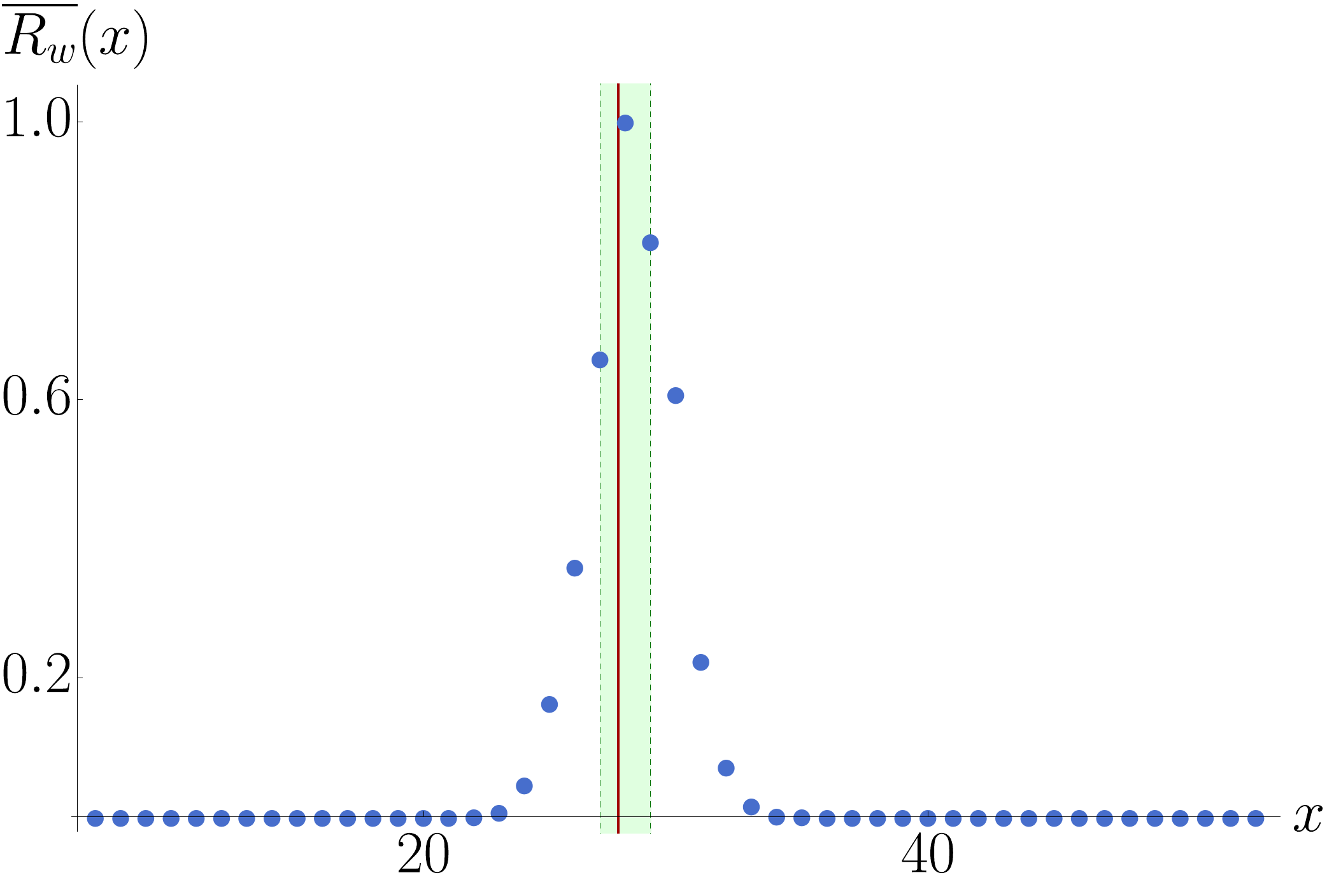}
  \end{subfigure}
  \hspace{3cm}
  \begin{subfigure}[c]{0.49\textwidth}
      \includegraphics[width=7cm]{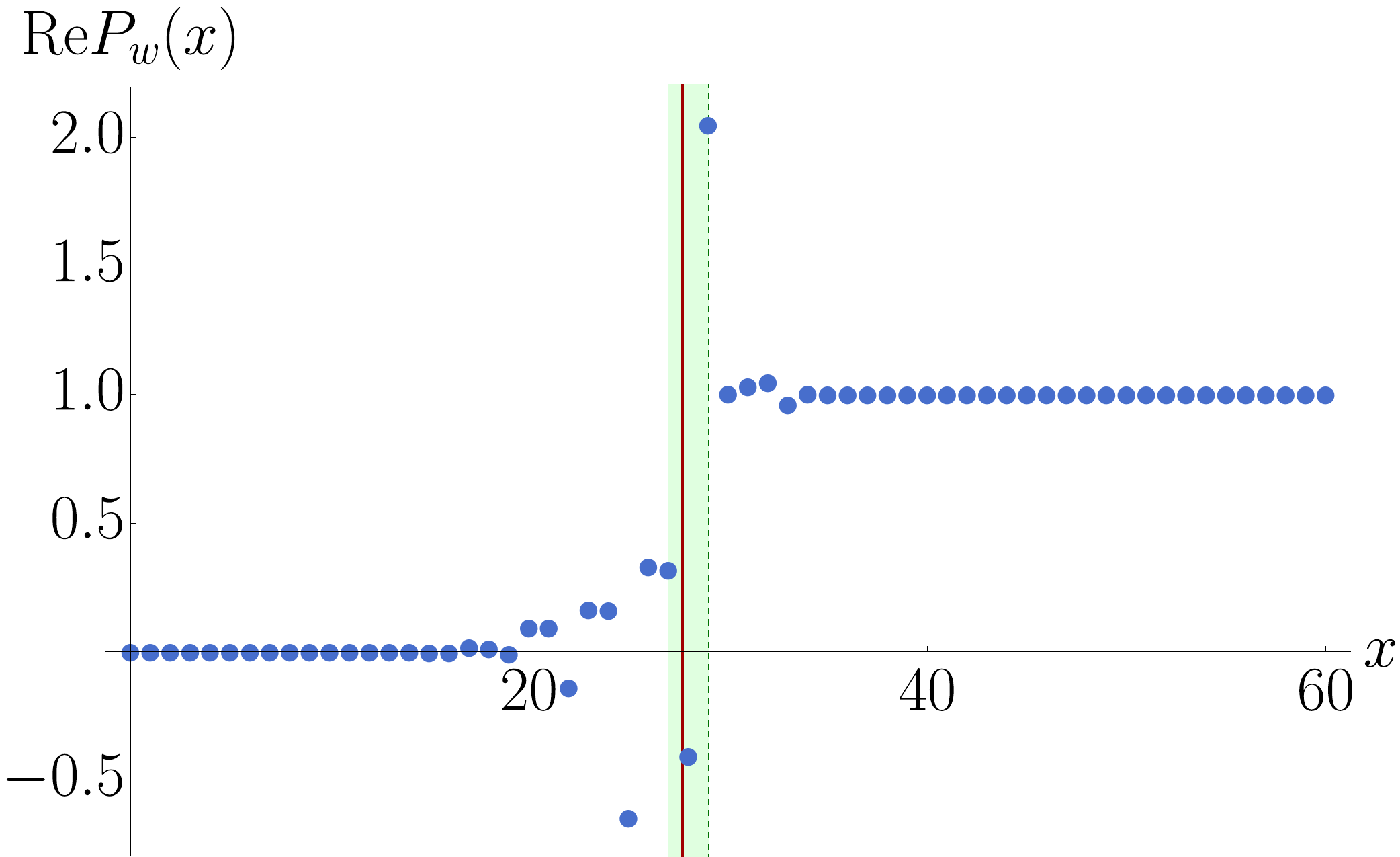}
  \end{subfigure}
  \caption{First curved configuration. Left: Averaged running sum $\overline{R_w(x)}$ with interval sizes $2$,$6$, and $10$. We normalize respect to the value of its maximum. Right: Real part of the partial sum $P_w(x)$ normalized respect the absolute value of the amplitude. In both plots, we mark with a red line the value of the geometrical area of the bulk triangle, and with a green band the values of the numerical estimate of the stationary phase point within one standard deviation.  }
  \label{fig:curvy1}
\end{figure}

\begin{figure}[H]
    \centering
    \begin{subfigure}[c]{4cm}
      \includegraphics[width=7cm]{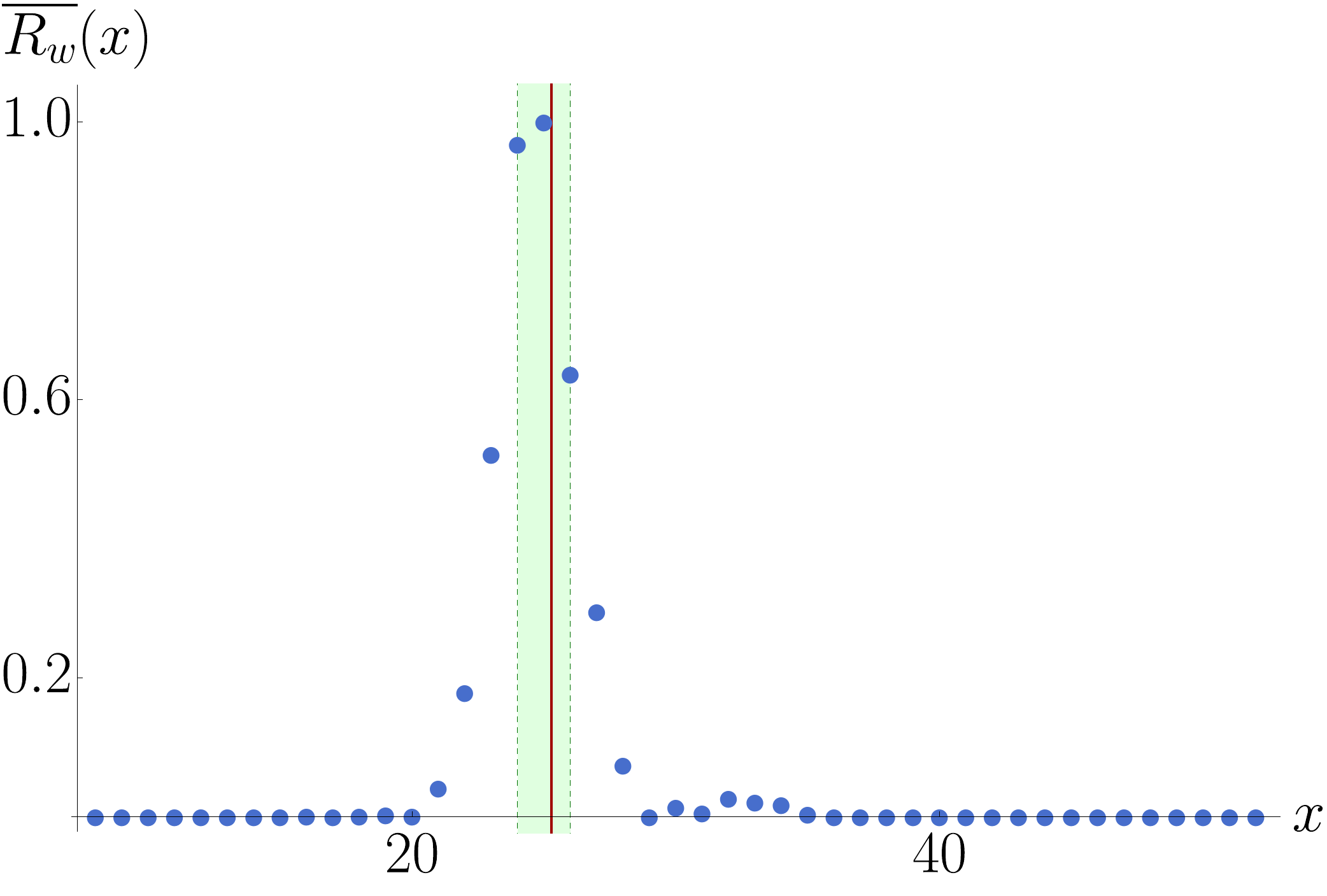}
    \end{subfigure}
    \hspace{3cm}
    \begin{subfigure}[c]{0.49\textwidth}
        \includegraphics[width=7cm]{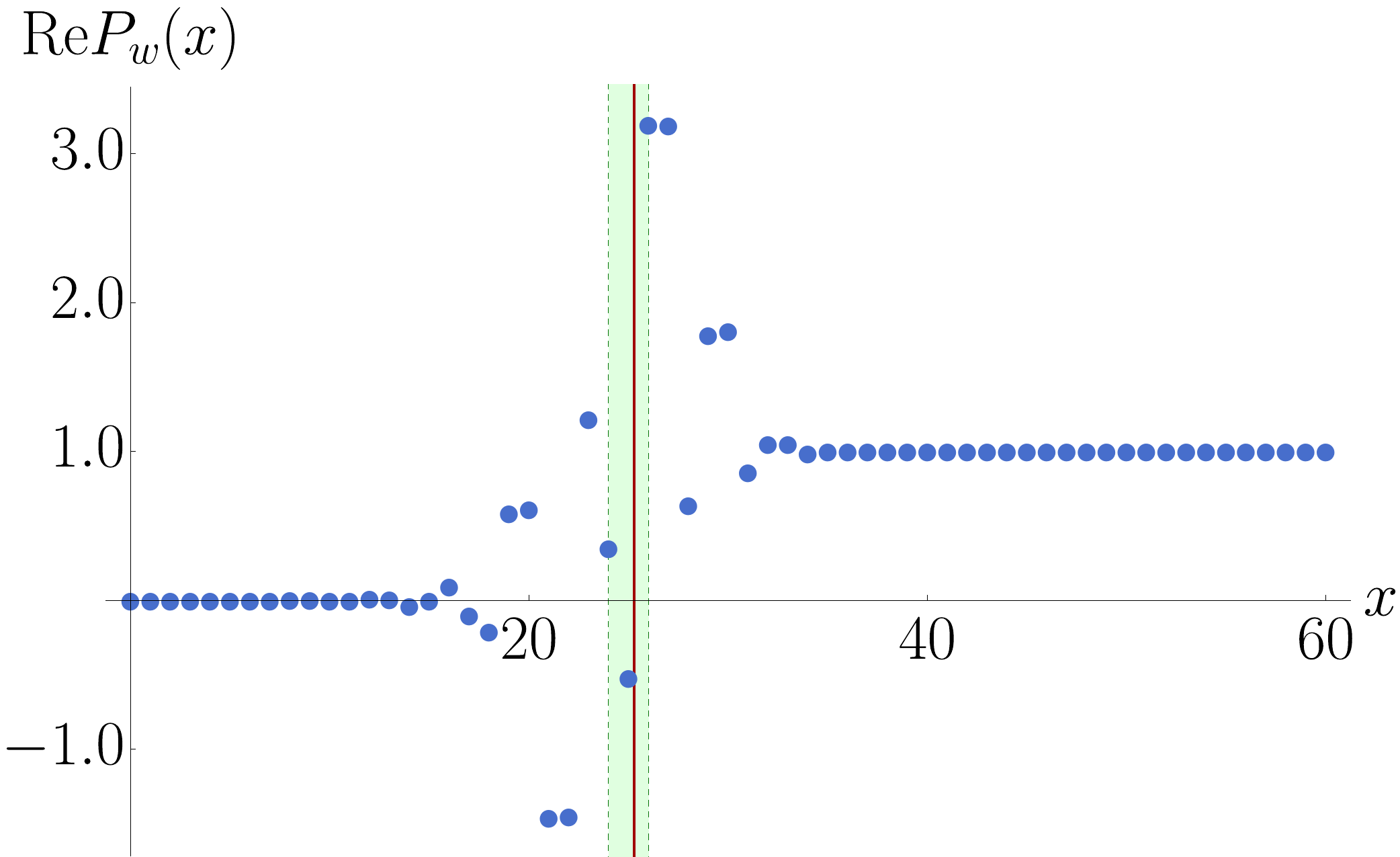}
    \end{subfigure}
    \caption{Second curved configuration. Left: Averaged running sum $\overline{R_w(x)}$ with interval sizes $2$,$6$, and $10$. We normalize respect to the value of its maximum. Right: Real part of the partial sum $P_w(x)$ normalized respect the absolute value of the amplitude. In both plots we mark with a red line the value of the geometrical area of the bulk triangle, and with a green band the values of the numerical estimate of the stationary phase point within one standard deviation. }
    \label{fig:curvy2}
  \end{figure}

We perform our numerical analysis also for boundary data corresponding to the two curved triangulations we described in \ref{sec:d3geometry} at scale $\lambda=20$. We report in Figure \ref{fig:curvy1} and \ref{fig:curvy2} the running sum and the partial sum for these configurations.

In both cases, we see a clear signature of the existence of a stationary phase point. 
For the first curved configuration we estimate the position of the stationary phase point for spin $x_{n_1}= 28 \pm 1$. It is compatible with the geometric area \eqref{eq:geoarea} of the curved triangulation $x_{g_1} =  1.39\ \lambda \approx 27.8$. Analogously, for the second curved configuration we find a stationary phase point at spin $x_{n_2}= 25 \pm 1$. Also in this case it is compatible with the expected geometric area $x_{g_2} = 1.26\ \lambda  \approx 25.2$.


\section{Flatness arguments applied to $W_{\Delta_3}$}
\label{sec:flatness}
In this section, we repeat the principal arguments for flatness of the EPRL model, on BF theory. To be specific we focus on the $W_{\dtre}$ amplitude \eqref{eq:d3ampl}. We sketch the various arguments to keep the exposition as clear as possible, and we refer to the original papers for more details. The arguments \cite{Conrady:2008mk,Bonzom:2009hw,Han:2013ina} are based on the large spins approximation  of the vertex amplitude. They approximate the integrals over the group using saddle point techniques, and then they look for stationary phase points in the sum over the bulk degrees of freedom. Inserting a resolution of the identity in terms of SU(2) coherent states in each internal edge the summand of the amplitude reduces to
\begin{align}
\label{eq:simplified}
w_\dtre(x,j_f,\vec{n}_{fa}) \propto  A_1(x,j_f,\vec{n}_{fa}) A_2(x,j_f,\vec{n}_{fa}) A_3(x,j_f,\vec{n}_{fa})
\end{align}
where $A_i(x,j_f,\vec{n}_{fa})$ are the $\{15j\}$ symbols contracted with coherent intertwiners.  The proportional symbol in \eqref{eq:simplified} indicates that we are omitting multiple dimensional factors and the integrals coming from the resolution of the identities we inserted. They both play no role in the calculation we illustrate in this section. 

If the boundary spins $j_f$ are large, we can safely assume that also the spin of the bulk face $x$ is large. The asymptotic expansion of the coherent $\{15j\}$ symbols was studied in \cite{Barrett:2009as,Dona:2017dvf}: if the boundary data of the coherent vertex forms the boundary a euclidean 4-simplex, the amplitude can be approximated with 
\begin{equation}
\label{eq:asymptotic}
A_i(x,j_f,\vec{n}_{fa})\approx N_{i}(x,j_f,\vec{n}_{fa}) \cos(S_i(x,j_f,\vec{n}_{fa}))
\end{equation}
where the factor $N_i$ contains numerical constants and the hessian, both of them are not relevant for the following considerations. The function $S_i = \sum_{f\in v_i} j_f \theta_{f,i} + x \theta_{x,i}$ is the Regge action for the 4-simplex: the sum runs over the triangles $f$ (faces) belonging to the ith 4-simplex (vertex) $v_i$, $j_f$ are the areas of the triangles and $\theta_{f,i}$ the $4$D dihedral angles around the triangle $f$ of the 4-simplex $i$. The dihedral angles can be reconstructed from orientation invariant scalar products between the normals $\vec{n}_{fa}$ using the spherical cosine law. In this analysis, we assume that vector geometries play no role or can be ignored by selecting the boundary data for the $\dtre$ amplitude appropriately.

In the large spin regime \eqref{eq:simplified} is proportional to 
\begin{align}
\label{eq:simplified2}
w_\dtre(x) \propto \cos(S_1(x))\cos(S_2(x))\cos(S_3(x))=&\, e^{i(S_1(x) + S_2(x) + S_3(x))}+e^{i(S_1(x) + S_2(x) - S_3(x))}+\\
& \nn \, e^{i(S_1(x) - S_2(x) + S_3(x))}+e^{i(-S_1(x) + S_2(x) + S_3(x))}+ c.c. \ ,
\end{align}
where we omitted the dependence on boundary spins $j_f$ and normals $\vec{n}_{fa}$ that we consider fixed from this point on. By linearity, we can search for stationary phase points of each of the eight terms in \eqref{eq:simplified2} independently and sum the results. We focus on the first term, the analysis of the others is similar. The derivative of the Regge action $S_i$ respect to the area $x$ is given by $\delta S_i(x)/ \delta x = \theta_{x,i}$. The stationary phase points are the solutions of the equation
\begin{equation}
\label{eq:statphaseeq}
\frac{\delta}{\delta x} (S_1(x) + S_2(x) + S_3(x)) = \theta_{x,1}+\theta_{x,2}+\theta_{x,3}=0 
\end{equation}
and this apparently implies that flat geometries dominate the summation over the bulk degrees of freedoms in the large spin limit. Notice that if we complete the analysis with the other terms in \eqref{eq:simplified2} we would obtain that the \emph{oriented} deficit angle is vanishing. In this section we disregard this additional but distinct complication, that was discussed in length in the \emph{proper vertex} literature \cite{Engle:2011un}.

The main counterargument to this analysis is that \eqref{eq:asymptotic} holds only if some constraints (closure and shape matching) are satisfied by the boundary data. Therefore, the variables are not independent and we cannot take the variation in \eqref{eq:statphaseeq} without including the constraints in the action.

Recently Oliveira \cite{Oliveira:2017osu} proposed a more accurate stationary phase point analysis taking into account also these constraints. They studied the spin foam transition amplitude on the $\dtre$ 2-complex in the euclidean EPRL model. The original paper claimed that in this setting the flatness problem was not present, however Kami\'nski and Engle \cite{EnglePrivateComm} corrected an oversight obtaining flatness once again. The analysis is based on the Malgrange preparation theorem. This approximation is valid in an entire neighborhood of a solution of the closure and shape matching constraints $x_0$. The effect on \eqref{eq:statphaseeq} is the addition of a purely imaginary term
\begin{equation}
\label{eq:statphaseeq2}
\frac{\delta}{\delta x} (S_1(x) + S_2(x) + S_3(x) + i \mu (x-x_0)^2 )= \theta_{x,1}+\theta_{x,2}+\theta_{x,3} + i\ 2\mu (x-x_0)=0 \ ,
\end{equation}
where $\mu$ depends on everything but $x$. To solve \eqref{eq:statphaseeq2} we need both the real and the imaginary part to vanish.
\footnote{The same conclusion is obtained in \cite{EnglePrivateComm} in a more rigorous way. Instead of looking for stationary phase points they perform the summation over $x$ using the Poisson resummation. The flatness condition is derived by requiring a non exponentially suppressed sum.} This is possible only if $x$ is a solution of the constraints $x=x_0$ and the flatness condition is satisfied, $\theta_{x,1}+\theta_{x,2}+\theta_{x,3}=0$.

Hellmann and Kami\'nski \cite{Hellmann:2013gva} suggested an argument for the flatness of the EPRL model based on a different strategy. First, they perform the summation over the bulk degrees of freedom and then they look for saddle points in the integrals over the holonomies. In the large boundary spins limit, the holonomies dominating the spin foam integrals are the ones contained in the wavefront set \cite{Hrmander2003} of the partition function. The wavefront set of $W_{\dtre}$ fixes the product of the group elements around its bulk face $g_x = \mathds{1}$. 
If we parametrize $g_x = \exp(i \Theta \, \vec{m} \cdot \vec{\sigma}/2)$ this condition imposes $\Theta=0$. Following \cite{Hellmann:2013gva} we can map
\footnote{For Regge boundary data, the coherent $\{15j\}$ symbol has two distinct saddle points. Up to gauge, the product of the two group elements on a wedge at the saddle point is characterized by $\pm$ the dihedral angle associated with that wedge. Each solution is responsible for one exponential forming the cosine in \eqref{eq:asymptotic}. If we take the saddle points of the three vertices with angles with the same sign, the angle of $g_x$ is $\pm$ the sum of dihedral angles around the shared triangle. With these two group elements, we use the map between $SU(2)\times SU(2)$ and $SO(4)$ to derive a non trivial $SO(4)$ geometrical holonomy, that is a rotation of angle $\Theta$ on the plane orthogonal to the shared triangle.} $g_x$ into the geometrical holonomy around the triangle shared by the three 4-simplices. In this case, $\Theta$ can be interpreted as $2\pi$ minus the deficit angle around the internal face, obtaining the flatness condition.


\section{Conclusion}
\label{sec:concl}

We presented a numerical exploration of the semiclassical limit of the $\dtre$ transition amplitude in the 4D BF spin foam model. We computed the transition amplitude with coherent boundary data corresponding to Regge triangulations, both flat and curved. We found a stationary phase point in the sum over the bulk spin with value compatible with the area of the dual triangle of the prescribed triangulation in all the three cases.

This was at first surprising. The \emph{flatness problem} argues that the partition function of the EPRL spin foam theory, in the semiclassical limit, is dominated by flat geometries. Usually, this is interpreted as an indication that the simplicity constraints, responsible for reducing a topological BF theory to General Relativity, are not imposed correctly. The same arguments, applied directly to SU(2) BF spin foam theory, indicate that curved geometries should be suppressed in the large boundary spins limit. This is in contradiction with our numerical result. 

What is the resolution of this apparent tension? In Section \ref{sec:flatness} we analyzed the most common flatness arguments, and we discussed where they might be misleading. In particular, some of them derive their results from factoring many-vertices amplitudes as the product of single-vertex amplitudes and then applying the known asymptotics expansion \cite{Barrett:2009gg} to each vertex. They conclude that contributions of flat geometries dominate the stationary point analysis. 

The same analysis can be applied to BF spin foam models as they have a similar asymptotic. The tension with our numerical calculation (which does \emph{not} involve any approximation) must come from the saddle-point analysis of the many-vertices factorization. We suggest that the angles appearing in this analysis should not be interpreted as the deficit angles computed with a spin connection, i.e. as deficit angles à la Regge. This is indeed the simplest explanation for the case under examination. The equations of motion impose flatness of the BF connection but do \emph{not} require the deficit angles computed with the piecewise flat metric (flat inside each 4-simplex) to vanish. The geometric deficit angles can assume any value, as indeed we observe in our numerical analysis.

In fact, the sums over the internal spins and the SU(2) integrals of the transition amplitude can be performed exactly, reducing it to a coherent $\{ 3n j \}$ symbol. The large spin asymptotic of general coherent SU(2) invariants is well understood \cite{Dona:2017dvf,DonaSpez}: the spin foam amplitude is not suppressed if the boundary data satisfy closure and shape matching constraints. The requirement of flat embedding in four-dimensional euclidean space is not necessary.

The same considerations may be partially applied to the EPRL spin foam model and variations, but the picture is less clear. Although these models have a semi-classical vertex expansion similar to BF models, the simplicity constraint are supposed to impose metricity on the spin connection.
Care is needed in interpreting the angles appearing in the asymptotic analysis of EPRL amplitudes as geometric deficit angles. Our results suggest that this interpretation and the current flatness claims are misleading. To make any claim of flatness, we need to have a clear understanding of the correct semiclassical spin foam dynamics of many-vertices triangulations and its relation with the associated Regge equations of motions.

In a recent paper \cite{Asante:2020qpa}, the authors explored the large spin limit of an effective spin foam model related to area-angle Regge calculus. They show that curved geometries are suppressed unless one considers a small Immirzi parameter in addition to large spins. They enrich their analysis with a numerical study confirming their findings. This calculation shows exactly the origin of the tension between our results and the various flatness arguments. These works share the common hypothesis that the asymptotic expression \eqref{eq:asymptotic} holds also for extended triangulations. We have preliminary indications \cite{me} that, while this expansion is correct for a single vertex, more care is needed in the presence of internal faces. All the flatness arguments become compatible with our numerical result if the angles in \eqref{eq:asymptotic} are not dihedral angles. What these arguments require to vanish is an angle related to the triangulation but that can not be interpreted as a pure deficit angle. 

To conclude, our result is twofold. On the positive side, we showed that the flatness claims of the EPRL spin foam models are misleading. In the context of BF spin foam theory, we found an explicit counter-example in which they are not valid. We argue that the same should be valid for the EPRL model. We believe that flatness arguments do not invalidate the EPRL model as a good candidate for the LQG dynamics. 
On the negative side, we remark that the semiclassical limit of spin foam theories is not yet well understood. We find evidence that the geometric Regge action does not emerge from the large spin limit of SU(2) BF theory. Our result suggests that this could also be the case for other spin foam models like the EPRL model. Therefore we should not content ourselves with the present status of semiclassical spin foam analyses. We still lack a complete study showing that the true Regge action with the correct geometric interpretation arises for more involved triangulations. In our opinion, this is an urgent problem: a better understanding of the relation between quantum and classical discrete geometry will be one of the most important steps forward to make spin foam theories a valuable candidate for quantum gravity. 

 We believe that numerics will play a central role in future studies of spin foams.  The numerical algorithm we described in this paper is powerful and can be applied to general amplitudes, with many vertices and internal faces, and different theories. We plan to use it to make a direct comparison of the large spin regime of BF and EPRL spin foam theories and finally provide an answer to this important question. 

\section{Addendum}

After the publication of this paper, we continued developing our numerical code. The completely new version \texttt{sl2cfoam-next} \cite{SLNEXT} improved on every aspect upon \texttt{sl2cfoam}, gaining several order of magnitudes in speed and efficiency. We expanded our case study, increasing the scale parameter to $\lambda=50$ for the three boundary data studied in this paper. We included a third curved triangulation\footnote{this case was suggested to us by Hongguang Liu in a discussion.} characterized by $\delta= 1.6$ and scale $\lambda=50$.

\begin{figure}
    \centering
    \begin{subfigure}[c]{0.495\textwidth}
        \includegraphics[width=\textwidth]{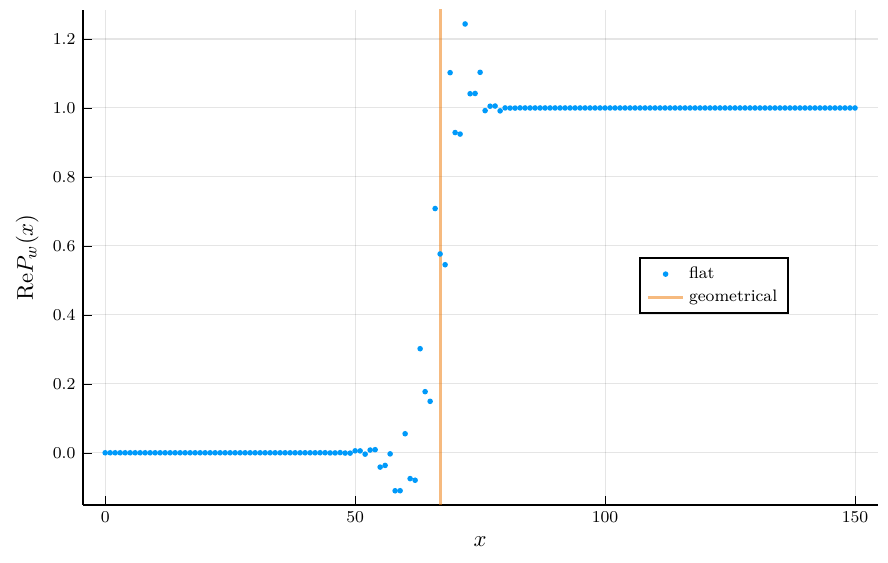}
    \end{subfigure}
    \begin{subfigure}[c]{0.495\textwidth}
        \includegraphics[width=\textwidth]{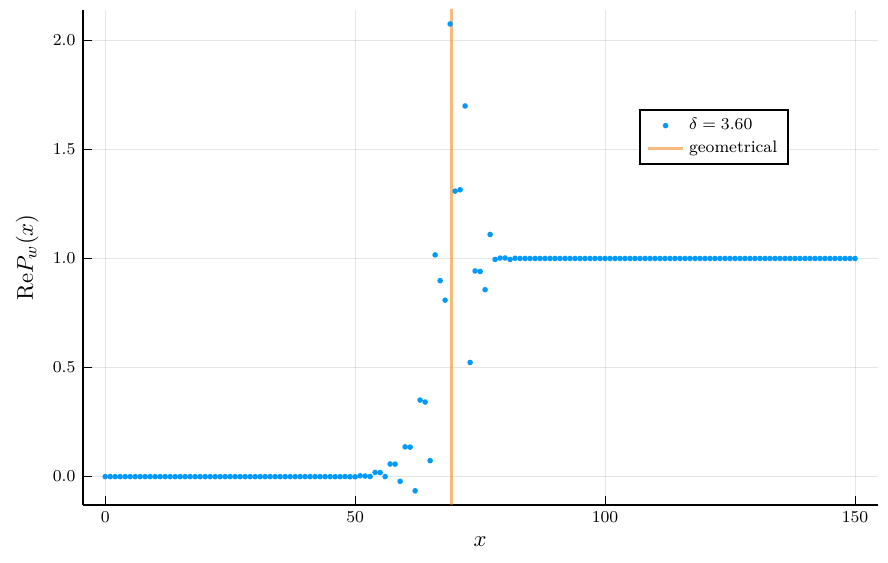}
    \end{subfigure}
    \begin{subfigure}[c]{0.495\textwidth}
        \includegraphics[width=\textwidth]{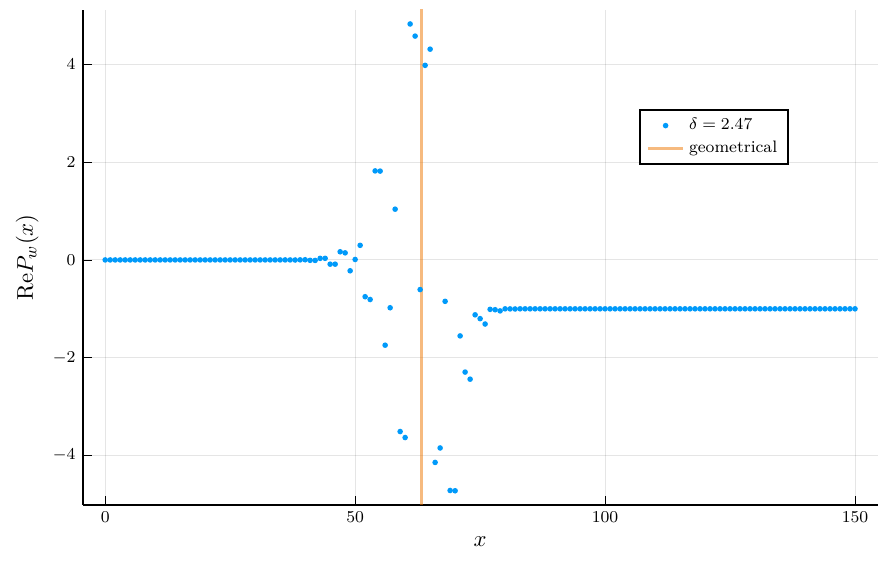}
    \end{subfigure}
    \begin{subfigure}[c]{0.495\textwidth}
        \includegraphics[width=\textwidth]{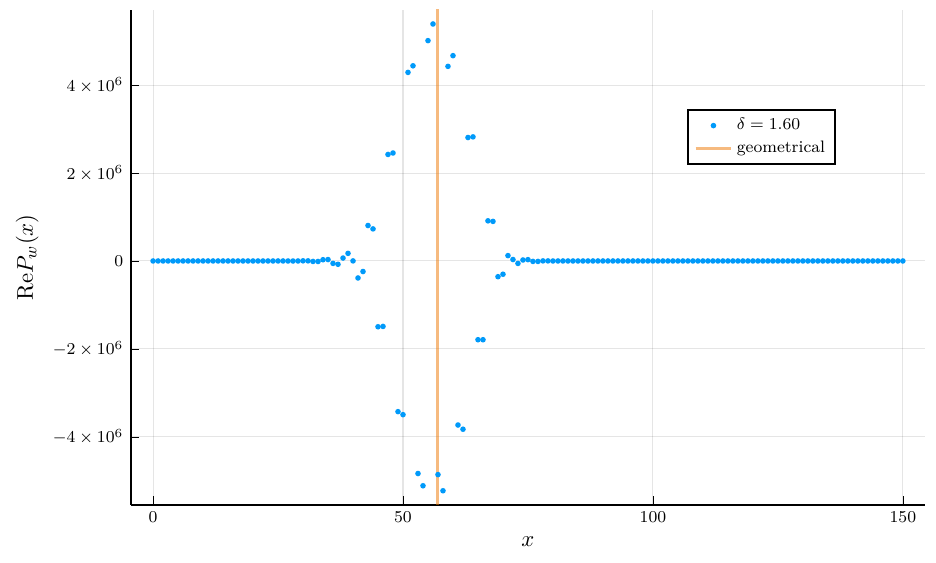}
    \end{subfigure}
    \caption{Real part of the partial sum $P_w(x)$ normalized with respect to the absolute value of the amplitude. In all plots we mark with a red line the value of the geometrical area of the bulk triangle. From top left to bottom right the four plots correspond to boundary data with $\lambda=50$ and $\delta=0$, $\delta=3.60$, $\delta=2.47$ and $\delta=1.6$.}
    \label{fig:psum50}
  \end{figure}

\begin{figure}
    \centering
    \begin{subfigure}[c]{0.495\textwidth}
        \includegraphics[width=\textwidth]{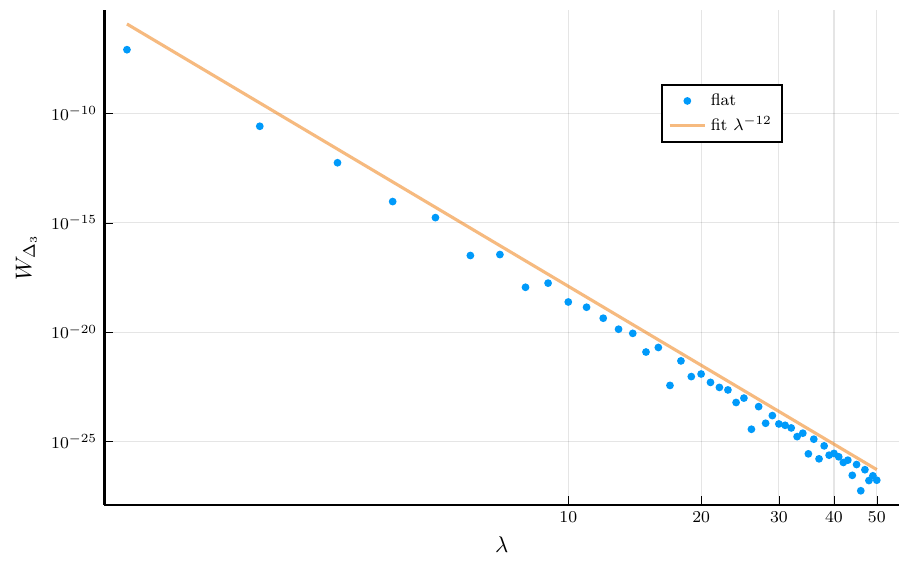}
    \end{subfigure}
    \begin{subfigure}[c]{0.495\textwidth}
        \includegraphics[width=\textwidth]{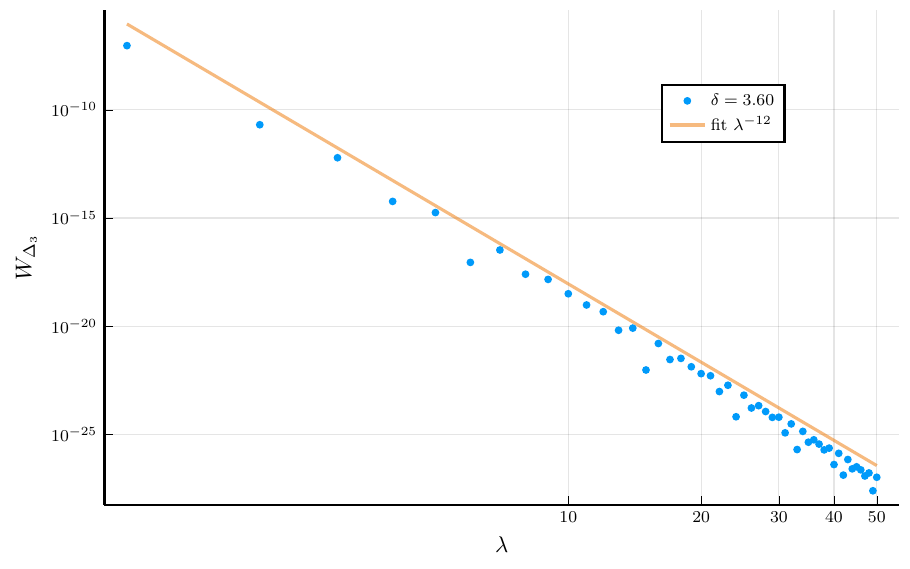}
    \end{subfigure}
    \begin{subfigure}[c]{0.495\textwidth}
        \includegraphics[width=\textwidth]{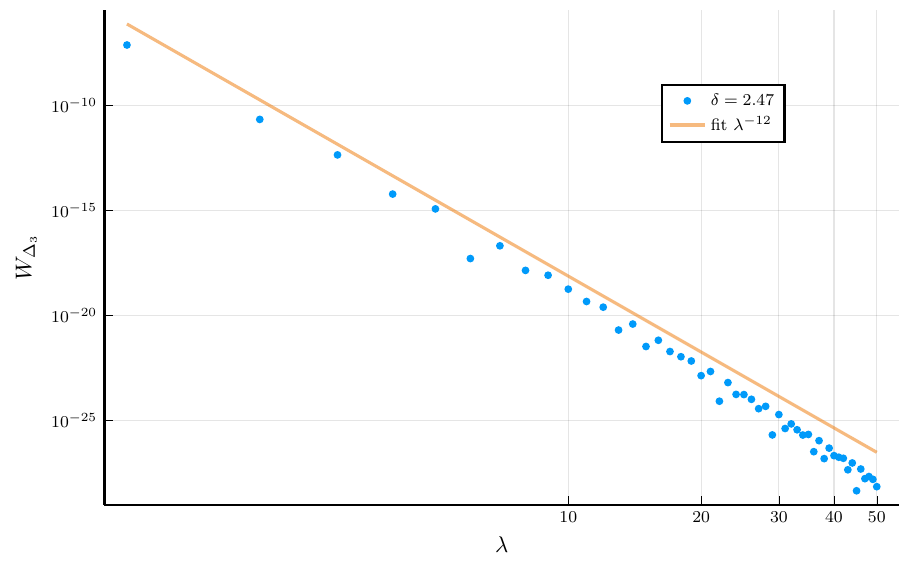}
    \end{subfigure}
    \begin{subfigure}[c]{0.495\textwidth}
        \includegraphics[width=\textwidth]{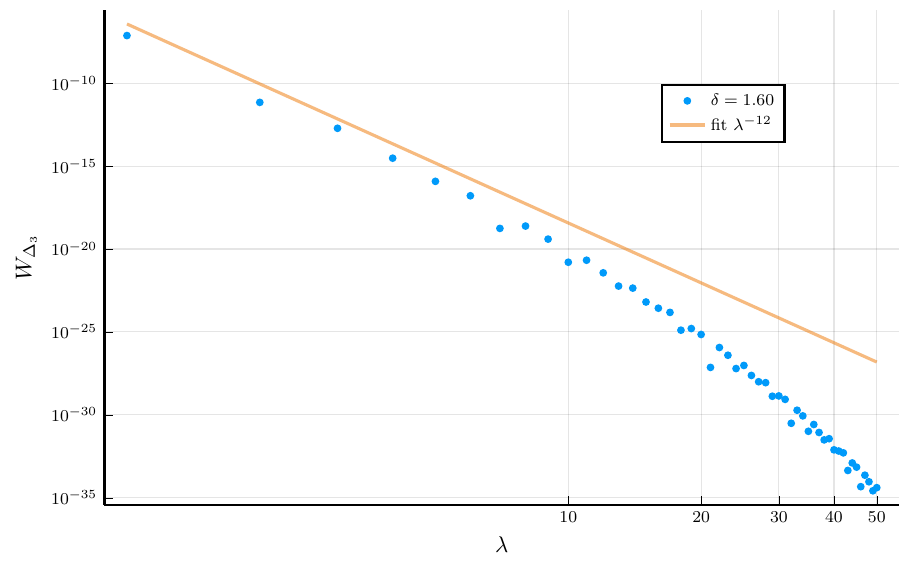}
    \end{subfigure}
    \caption{Log-Log plot of the absolute value of the $W_\dtre$ amplitude as a function of the scale $\lambda$. From top left to bottom right the four plots correspond to boundary data with $\lambda=40$ and $\delta=0$, $\delta=3.60$, $\delta=2.47$ and $\delta=1.6$. We also plot the power-law $\lambda^{-12}$ for comparison. The scaling of the flat configuration is compatible with a power-law, while the scaling of the configuration with $\delta=1.6$ is exponentially suppressed. With the other two boundary data it is difficult to distinguish a power-law scaling from an exponential suppression.}
    \label{fig:scal50}
  \end{figure}

While the stationary phase point for flat boundary data is stable under rescaling, for curved boundary data we can infer a new trend. The characteristic oscillations of the partial sum around the ``stationary phase point'' can be many orders of magnitude greater than the value of the total amplitude, see Figure \ref{fig:psum50}. This is evident for the new curved configuration we included here. We can interpret this as a signal of an exponential suppression of the full amplitude. We also studied the scaling behavior of the full amplitudes for the four boundary data, see Figure \ref{fig:scal50}.

While the exponential suppression is particularly marked for the new curved boundary data, it is barely noticeable for the curved boundary data previously considered. In light of the recent results \cite{ADDENDUM} we can interpret this difference as larger and smaller exponential suppression coefficients due to the accidental curvature constraint. Similar results can also be obtained for the EPRL model and will be published soon \cite{SLNEXT}.

We have to reconsider our conclusions. Our numerical analysis looks for the bulk spins that dominate the summation. We interpreted these points as stationary phase points. However, this interpretation seems to be inaccurate. We do not have numerical access to the action that enters the asymptotic analysis. We can infer the configurations that dominate the summation over bulk spins, but since these cannot be true stationary phase points in a discrete summation, we cannot derive from our analysis the exact scaling and the eventual exponential suppression of the resummed amplitude. We can find hints of the exponential suppression by noticing the presence of very large oscillations compared to the final value of the amplitude, but still the exponential suppression might be so slow that it could not be seen in the range of spins we have access to numerically. We conclude that, although still both flat and curved geometries dominate the sum over bulk degrees of freedom, the accidental curvature constraint in the curved case is present and results in an exponential suppression of the total amplitude in the large spin limit.

\section{Acknowledgments}
The work of P.D. is partially supported by the grant 2018-190485 (5881) of the Foundational Questions Institute and the Fetzer Franklin Fund. We thank Lorenzo Bosi for his technical support with the CPT servers. F.G. acknowledges the Centre de Calcul Intensif d'Aix-Marseille for granting access to its high performance computing resources. We thank Carlo Rovelli for very useful discussions and for encouraging us in writing this paper. Special thanks go to Simone Speziale, for motivating us not to give up on this project, helping us with countless inputs and ideas, while taking care of another new little piece of his life. We thank Muxin Han, Hongguang Liu, and Jonathan Engle for suggesting us to expand our analysis and for pointing out the problem in our conclusions.

\begin{appendices}

  \section{Convention for the $\{15j\}$ symbol}
  \label{app:conventions}
The $\{15j\}$ symbol we use in this work is the irreducible $\{15j\}$ symbol of first type, following the convention of \cite{Yutsis:1962vcy}. The definition and its graphical representation is the following
  \begin{align}
  \label{eq:15jconv}
  \raisebox{-15mm}{ \includegraphics[width=3.5cm]{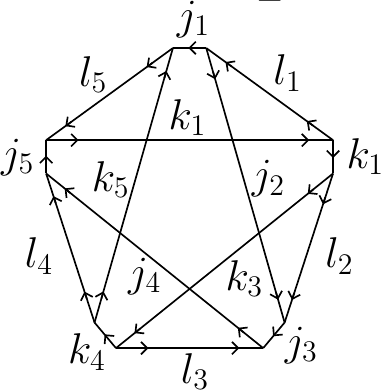}} &= \left \{ \begin{array}{ccccc} j_1 & j_2 & j_3 & j_4 & j_5 \\  l_1 & l_2 & l_3 & l_4 & l_5 \\ k_1 & k_2 & k_3 & k_4 & k_5 \end{array}\right \}  =\\
  &=  (-1)^{\sum_{i=1}^5 j_i + l_i +k_i} \sum_s (2 s +1) \Wsix{j_1}{k_1}{s}{k_2}{j_2}{l_1} \Wsix{j_2}{k_2}{s}{k_3}{j_3}{l_2} \\ & \hspace{3cm }\times \Wsix{j_3}{k_3}{s}{k_4}{j_4}{l_3} \Wsix{j_4}{k_4}{s}{k_5}{j_5}{l_4} \Wsix{j_5}{k_5}{s}{j_1}{k_1}{l_5}  \ .
  \end{align}

  \section{Coherent states} 
  \label{app:coherentstates}

  Coherent states are a fundamental ingredient to study the semiclassical behavior of a spin foam amplitude. We briefly review the conventions used in this paper. For more details we refer to \cite{Perelomov1986} or to other numerical works \cite{Dona:2017dvf} and \cite{Dona:2018nev}.
  A SU(2) coherent state $\ket{j, \vec{n}}$ in the irreducible representation of spin $j$ is given by the action on the lowest weight state $\ket{j, - j}$ of a SU(2) group element $g_n$ corresponding to the rotation that transforms the unitary vector $\vec{z} = (1,0,0) $ into $\vec{n} = (\sin{\Theta}\cos{\Phi}, \sin\Theta\sin\Phi,\cos{\Theta})$. This coherent state minimizes the uncertainty in the two directions orthogonal to $\vec{n}$. The Euler angles of group element $g_n$ are related to the angles $\Theta$ and $\Phi$ by
  \begin{equation}
    g_n = e^{-i\Phi \frac{\sigma_z}{2}} e^{-i\Theta \frac{\sigma_y}{2}} e^{i\Phi \frac{\sigma_z}{2}} \ ,
  \end{equation}
where we expressed the SU(2) generators $i \vec{\sigma}/2$ in terms of the Pauli matrices. The scalar product between a coherent state and a state in the magnetic basis is given by
  \begin{equation}
   \bra{j, m}j, \vec n \ra =  \bra{j, m} D^{(j)}(g_n) \ket{j,\vec{n}} = D^{(j)}_{m-j}(g_n)=D^{(j)}_{m-j}(\Phi,\Theta,-\Phi) \ ,
  \end{equation}
where $D^{(j)}_{m-j}(\Phi,\Theta,-\Phi)$ is the Wigner matrix element of $g_n$ parametrized in terms of its Euler angles. These conventions are used by \texttt{sl2cfoam} and Wolfram Mathematica.

 SU(2) coherent states form an overcomplete basis in the irreducible representation of spin $j$ space 

  \begin{equation}
   \mathds{1}^{(j)} =  (2j+1)\int_{S^2} \mathrm{d} \vec{n} \ket{j,\vec{n}}\bra{j,\vec{n}}  \ ,
  \end{equation}
where $\mathrm{d} \vec{n}$ is the normalized measure on the sphere.

A 4-valent intertwiner in the recoupling basis is given by 
\begin{equation}
\ket{j_i ; j_{12}} = \sum_{m_1} \Wfour{ j_1}{ j_2}{ j_3} {j_4}{m_1}{m_2}{m_3}{m_4}{j_{12}} \ket{j_1 m_1}\ket{j_2 m_2}\ket{j_3 m_3}\ket{j_4 m_4}
\end{equation}
and the $(4jm)$ symbol is given by the contraction
\begin{equation}
\Wfour{ j_1}{ j_2}{ j_3} {j_4}{m_1}{m_2}{m_3}{m_4}{j_{12}} = \sum_{m} (-1)^{j_{12}-m} \Wthree{ j_1}{ j_2}{ j_{12}} {m_1}{m_2}{m} \Wthree{ j_{12}}{ j_3}{ j_4} {-m}{m_3}{m_4}\ .
\end{equation}
Finally we can define a coherent tetrahedron as the projection on the intertwiner space of the tensor product of four SU(2) coherent states. Each one of them is associated to a face of the tetrahedron with area $j_i$ and normal $\vec{n}_i$
\begin{align}
\label{coeffCS}
c_{j_{12}}(\vec{n}_i) :& = \bra{j_i;j_{12}} j_1, \vec n_1 ; j_2,\vec{n}_2; j_3,\vec{n}_3; j_4,\vec{n}_4 \ra \\ & = \Wfour{ j_1}{ j_2}{ j_3} {j_4}{m_1}{m_2}{m_3}{m_4}{j_{12}}  D^{(j_1)}_{m_1-j_1}(g_{n_1}) D^{(j_2)}_{m_2-j_2}(g_{n_2}) D^{(j_3)}_{m_3-j_3}(g_{n_3}) D^{(j_4)}_{m_4-j_4}(g_{n_4}) \ .
\end{align}
The graphical representation of an SU(2) coherent state is given by a line ending on a little circle while the coherent states coefficients are represented by
\begin{equation}
\ket{j,\vec{n}} = \raisebox{-1.8mm}{\includegraphics[width=0.25cm]{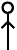}} \qquad c_{j_{12}}(\vec n_i) =   \raisebox{-5mm}{\includegraphics[width=2.8cm]{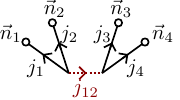}} \ .
\end{equation}

\section{Derivation of the $\Delta_3$ spin foam amplitude} 
\label{app:derivation}

In this appendix we derive the formula \eqref{eq:d3ampl} for the spin foam transition amplitude associated to the $\Delta_3$ triangulation starting from its integral representation. We use the SU(2) graphical calculus to represent the amplitude as
\begin{equation}
\label{eq:d3ampfull}
  W_{\Delta_3} (j_f,\vec{n}_{fa}) = \sum_x (2x + 1) \raisebox{-30 mm}{ \includegraphics[width=8cm]{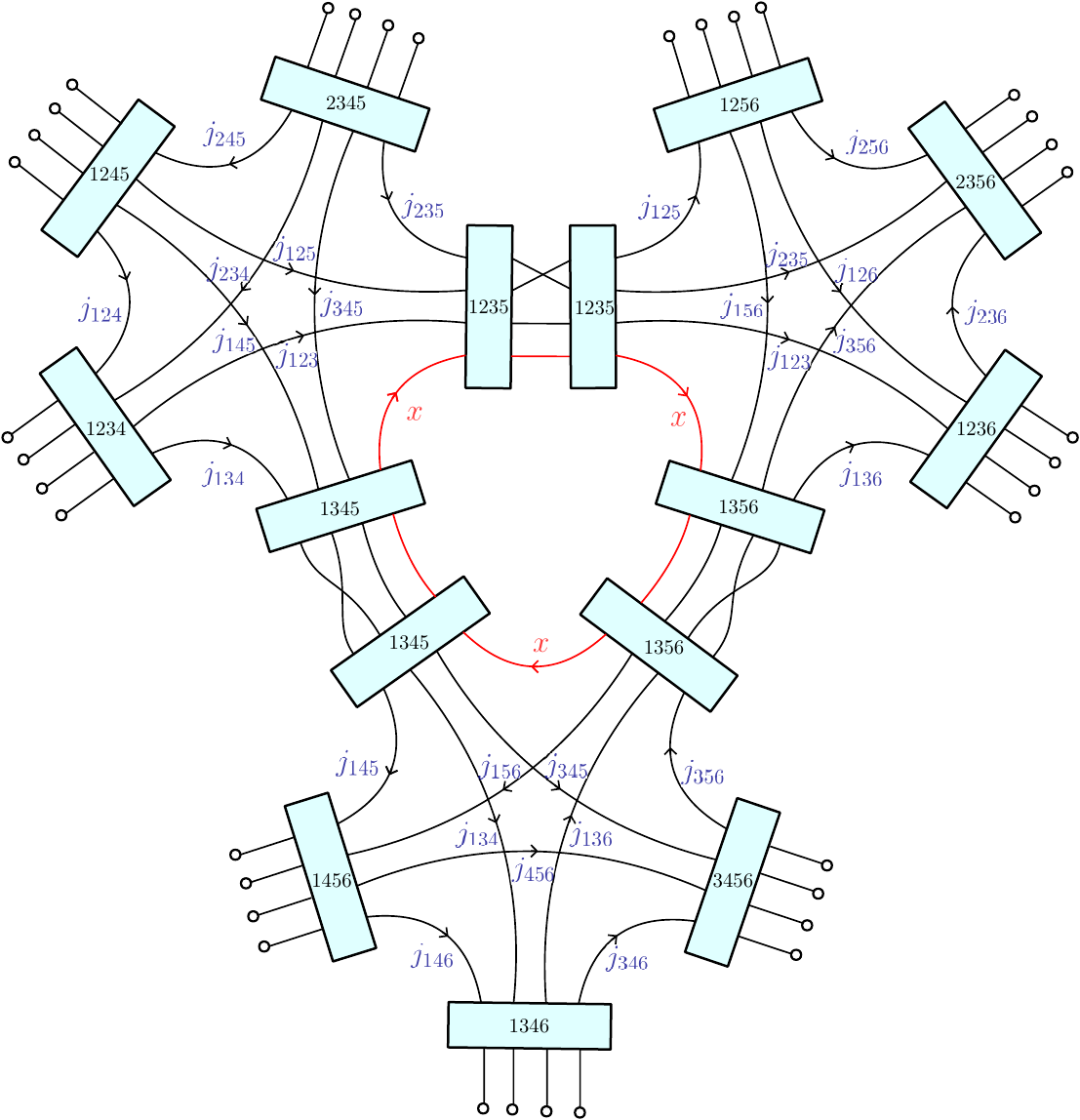}}
\end{equation}
where we picked a conventional orientation of the faces that are capped by coherent states on both ends, represented by empty circles. We label the 4-simplices using the five points, the edges, dual to a tetrahedron, are labeled by four points and the faces, dual to triangles, are labeled by three points. We highlight in red the internal face $x$. 

We perform the integrations over $SU(2)$ and we get
\begin{align}
  W_{\Delta_3} (j_f,\vec{n}_{fa})   &=  \sum_x (2 x +1) \sum_{k_b} \left( \prod_b (2 k_b + 1)^2 \right) \sum_{i_a} \left( \prod_a (2 i_a + 1) c_{i_a}(\vec{n}_{fa}) \right) \times \\ & \raisebox{-20 mm}{ \includegraphics[width=9cm]{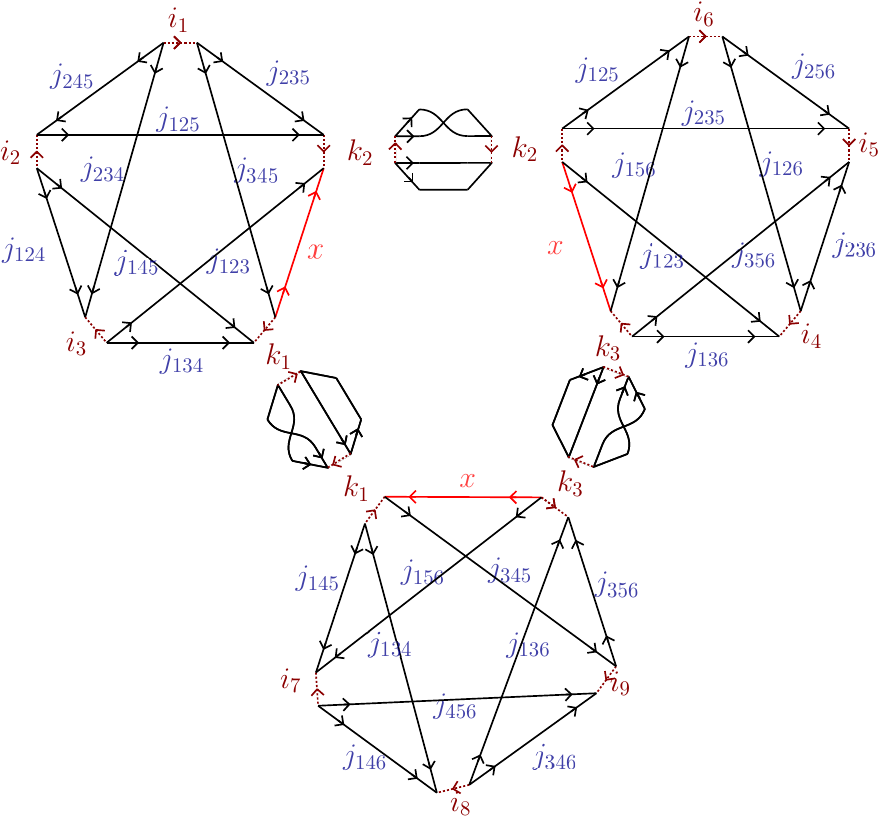}} \ .
\end{align}
To keep the picture as simple as possible we opted for not using the graphical representation of the coherent tetrahedra $c_{i_a}(\vec{n}_{fa})$ that we factorize. The ``dipole'' like diagrams contribute with a phase and the inverse dimension of the intertwiner space
\begin{equation}
 \label{phase}
 \raisebox{-6.2 mm}{ \includegraphics[width=2cm]{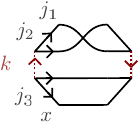}} = \frac{(-1)^{k + j_1 +j_2}} {2k + 1} \ .
\end{equation}
We invert the orientation of some internal lines to conform with \eqref{eq:15jconv} and obtain the $\Delta_3$ transition amplitude in terms of three $\{15j\}$ symbols
\begin{align}
  W_{\Delta_3} (j_f,\vec{n}_{fa})   &= (-1)^{\chi} \sum_x (-1)^x(2 x +1) \sum_{k_b} \left( \prod_b (-1)^{k_b}(2 k_b + 1)\right) \sum_{i_a} \left( \prod_a (2 i_a + 1) c_{i_a}(\vec{n}_{fa}) \right) \\ & \times \raisebox{-20 mm}{ \includegraphics[width=12cm]{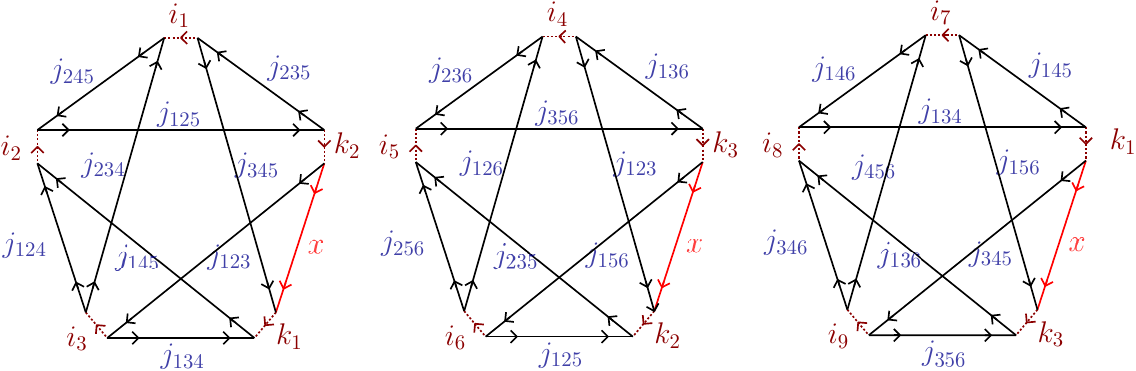}}
\end{align}
with $\chi = 2 (j_{123} + j_{234}+ j_{124}+ j_{134}+j_{456}+j_{156}+j_{346}+j_{356}+j_{235}) + j_{123} + j_{345} + j_{156}$. This phase is obtained from the contributions of \eqref{phase} and $(-1)^{2j}$ required to invert the orientation of an internal line of spin $j$. We finally derive \eqref{eq:d3ampl}.

\end{appendices}


\end{document}